\documentclass[onecolumn, 12pt]{IEEEtran}
\usepackage{amsmath,amsfonts,amssymb,enumerate,cite}
\usepackage[final]{lscape,graphicx,epsfig}
\usepackage{cite}

\DeclareMathOperator*{\argmin}{argmin}

\newtheorem{thm}{Theorem}

\title{On Inequalities Relating the Characteristic Function and Fisher Information}

\author{Cihan Tepedelenlio\u{g}lu, {\em Member,
IEEE}, Mahesh K. Banavar, {\em Student Member, IEEE} and Andreas
Spanias, {\em Fellow, IEEE} \thanks{The work in this paper is
supported by the SenSIP Center, Arizona State University. C.
Tepedelenlio\u{g}lu, M.K. Banavar and A. Spanias are with the SenSIP
Center, School of Electrical, Computer and Energy Engineering,
Fulton Schools of Engineering, Arizona State University, Tempe, AZ
85287. USA. (Email: \{cihan, maheshkb, spanias\}@asu.edu). }}

\date{}

\begin{document}

\maketitle
\bibliographystyle{IEEEtran}

\begin{abstract}
A relationship between the Fisher information and the characteristic
function is established with the help of two inequalities. A necessary and sufficient condition for equality is found. These results are used to determine the asymptotic efficiency of a distributed estimation algorithm that uses constant modulus transmissions over Gaussian multiple access channels. The loss in efficiency of the distributed estimation scheme relative to the centralized approach is quantified for different sensing noise distributions. It is shown that the distributed estimator does not incur an efficiency loss if and only if the sensing noise distribution is Gaussian. 
\end{abstract}

\begin{IEEEkeywords}
Fisher information, characteristic function, asymptotic efficiency, wireless sensor network, distributed estimation
\end{IEEEkeywords}

\section{Introduction}
\label{sec:introduction}

We investigate the relationship between the Fisher
information about a location parameter and the characteristic function of the additive noise by providing a new
derivation for two inequalities that involve the Fisher information and the
characteristic function. These inequalities were originally derived using a different approach and applied in a quantum physics setting to estimate the survival probability of a quantum state in \cite{chinesephysics}. Conditions for equality are also delineated herein for
the first time in the literature, and used to investigate the asymptotic efficiency of a distributed estimation scheme over a Gaussian multiple-access channel. 

\section{The Inequalities}
\label{sec:asv_eff}

Consider a model where a deterministic location parameter, $\theta$, is related to
observations $x_{l} = \theta + \eta_{l}$, $l=1, \dots, L$, where $\eta_{l}$ are iid and
real-valued random variables. Let the characteristic function of $\eta_{l}$ be
$\varphi(\omega) \mathop{:=} E [e^{j \omega \eta_{l}}]$ and let the Fisher information be defined as \cite{Cover91, zamir}
\begin{equation}
I(\eta)  \mathop{:=} \int_{-\infty}^{\infty} \frac{\left[ p'(x)
\right]^{2}} {p(x)} dx < \infty, \label{eqn:fisher_info_defn}
\end{equation}
where $p(x)$ is the pdf of $\eta_{l}$, assumed to be continuously differentiable, and with support $(-\infty, \infty)$. Note that $I(\eta)$ is the Fisher information in $x_{l}$ about $\theta$, and is a deterministic value which does
not depend on $\theta$. In the following, $\eta$ denotes a random variable with the same distribution as any $\eta_{l}$. 

We present the following theorem, which
provides two bounds involving $I(\eta)$ and $\varphi(\omega)$. It
was proved first in \cite{chinesephysics} using the Cram\'{e}r-Rao inequality. We provide an alternate
proof which also delineates the condition for equality for the first
time in the literature. The condition for equality will be central in Section \ref{sec:model} to establish necessary and sufficient conditions for the asymptotic efficiency of a distributed estimation algorithm over a Gaussian multiple-access channel. 

\begin{thm}
Let $\varphi_{R}(\omega)$ and $\varphi_{I}(\omega)$ be the
real and the imaginary parts of $\varphi(\omega)$, respectively. 
We have
\begin{align}
\omega^{2} \varphi_{I}^{2}(\omega) &\leq I(\eta) \left[ \frac{1}{2}
\left[ 1 + \varphi_{R} (\omega) \right] - \varphi_{R}^{2}(\omega)
\right], \label{eqn:lem_ineq_imag} \\
\omega^{2}
\varphi_{R}^{2}(\omega) &\leq I(\eta) \left[ \frac{1}{2} \left[ 1 -
\varphi_{R} (\omega) \right] - \varphi_{I}^{2}(\omega) \right],
\label{eqn:lem_ineq_real}
\end{align}
with equality in both (\ref{eqn:lem_ineq_imag}) and
(\ref{eqn:lem_ineq_real}) if and only if $\omega = 0$.
\label{lemma:bounds_fisher_info}
\end{thm}

\begin{IEEEproof}
Let $s(x) \mathop{:=} p'(x)/p(x)$ be the score function, where we recall that $p(x)$ is the
pdf of $\eta_{l}$. Let $g(x)$ be a differentiable function satisfying
$\lim_{x\to\pm\infty} g(x) p(x) = 0$. Using Stein's identity \cite[Lemma
1.18]{Johnson2004}, we have
\begin{equation}
E \left[ g(\eta) s(\eta) \right] = -E \left[ g'(\eta) \right].
\label{eqn:lemproof_exp_der}
\end{equation}
Applying the Cauchy-Schwarz inequality yields
\begin{equation}
E^{2}[g'(n)] \leq I(\eta) E[g^{2}(\eta)],
\label{eqn:lemproof_cuachy_schwarz}
\end{equation}
with equality if and only if $s(x)=\alpha g(x)$ for some $\alpha$
and all $x$. By substituting $g_{1}(x) \mathop{:=} \cos(\omega x) -
\varphi_{R}(\omega)$ for $g(x)$ in (\ref{eqn:lemproof_cuachy_schwarz}), equation (\ref{eqn:lem_ineq_imag}) is obtained. Similarly, 
$g_{2}(x) \mathop{:=} \sin(\omega x) - \varphi_{I}(\omega)$ substituted for $g(x)$ yields equation
(\ref{eqn:lem_ineq_real}).

To examine when equality occurs, first note that if $\omega=0$,
since $\varphi_{R}(0) = 1$ and $\varphi_{I}(0) = 0$, equations
(\ref{eqn:lem_ineq_imag}) and (\ref{eqn:lem_ineq_real}) become
equalities. Conversely, consider $\omega \neq 0$. The equality
condition for (\ref{eqn:lem_ineq_real}) is $s(x) = \alpha g_{2}(x)$,
which yields the first order differential equation
\begin{equation}
\frac{p'(x)} {p(x)} = \alpha \left[ \sin(\omega x) -
\varphi_{I}(\omega) \right], \label{eqn:lemproof_diff_eqn}
\end{equation}
which must provide a solution satisfying $p(x)\geq 0$ and
$\int_{-\infty}^{\infty} p(x) dx = 1$. The solution to
(\ref{eqn:lemproof_diff_eqn}) is of the form $p(x) = C e^{-\alpha x
\varphi_{I}(\omega)} e^{-\frac{\alpha} {\omega} \cos(\omega x)}$,
which is unbounded as $x\to -\infty$ when $\varphi_{I}(\omega) \neq 0$, and periodic
when $\varphi_{I}(\omega) = 0$. In either case, $\int_{-\infty}^{\infty} p(x) dx = 1$ is not possible. 
This shows that there is no pdf satisfying
(\ref{eqn:lemproof_diff_eqn}) when $\omega \neq 0$, and therefore,
equality in (\ref{eqn:lem_ineq_real}) cannot be attained for
$\omega \neq 0$. The same conclusion can be drawn about
equation (\ref{eqn:lem_ineq_imag}), using a similar argument with $s(x) = \alpha g_{1}(x)$.
\end{IEEEproof}

\section{Application to Distributed Estimation}
\label{sec:model}

\begin{figure}[tb]
\begin{minipage}[b]{1.0\linewidth}
  \centering
  \centerline{\epsfig{figure=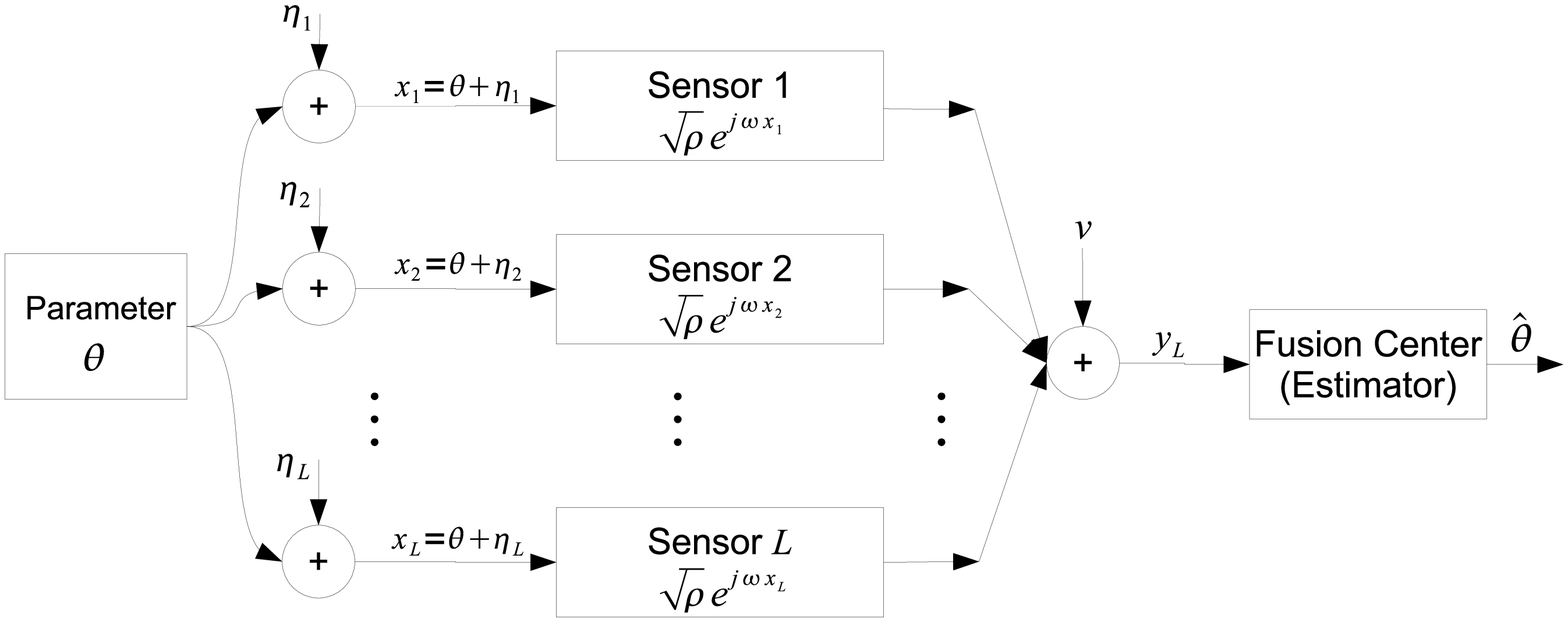,width=11cm,height=5cm}}
\end{minipage}
\caption{System model: Wireless sensor network. The estimator is located at the fusion center.} \label{Fig:problem_setup}
\end{figure}

A sensor network, illustrated in Figure \ref{Fig:problem_setup},
consisting of $L$ sensors is considered. The value, $x_{l}$, observed at the
$l^{th}$ sensor is
\begin{equation}
x_{l} = \theta + \eta_l \label{eqn:obs_eqn}
\end{equation}
for $l = 1,...,L$, where $\theta$ is a deterministic,
real-valued, unknown parameter in a bounded interval of known length, $[0, \theta_{R}]$, where
$\theta_{R}<\infty$, and $\eta_{l}$ are iid real-valued random variables. 
We will assume that $\eta_{l}$
has zero mean and variance $\sigma_{\eta}^{2}$, when the mean and variance exist.  Due to constraints in the transmit power, we consider a scheme where the
$l^{th}$ sensor transmits its measurement, $x_{l}$, using a constant modulus base-band equivalent
signal, $\sqrt{\rho} e^{j \omega x_{l}}$, over a Gaussian multiple
access channel so that the received signal at the fusion center is given by
\begin{equation}
y_{L} = \sqrt{\rho} \sum_{l=1}^{L} e^{j \omega x_{l}} + \nu,
\label{eqn:yn}
\end{equation}
where the transmitted signal at each sensor has
per-sensor power of $\rho$, $\omega\in(0,2\pi /\theta_{R}]$ is a
design parameter to be optimized, and $\nu \sim \mathcal{CN}(0,\sigma_{\nu}^{2})$
is independent of $\{\eta_{l}\}_{l=1}^{L}$. Note that the
restriction $\omega\in(0,2\pi /\theta_{R}]$ is necessary even in the
absence of sensing and channel noise ($y_{L} = \sqrt{\rho} e^{j
\omega \theta}$) to uniquely determine $\theta$ from $y_{L}$.

In a centralized problem, $\theta$ is estimated from $\{x_{l}\}_{l=1}^{L}$. The Cram\'{e}r-Rao bound is the well known benchmark on the variance of unbiased estimators with finite samples and is proportional to $[I(\eta)]^{-1}$ \cite[pp. 120]{Lehmann1998}. For large $L$, the asymptotic variance is an appropriate performance metric. Under certain regularity conditions, the benchmark on the asymptotic variance is given by $[I(\eta)]^{-1}$ \cite[pp. 439]{Lehmann1998}. Hence, the Fisher information has a central role to play in establishing benchmarks for the estimation of a location parameter for centralized estimation problems which address estimators of $\theta$ based on $\{x_{l}\}_{l=1}^{L}$. 

For the distributed setting, based on (\ref{eqn:yn}), the estimators of $\theta$ rely on $y_{L}$. The desire to have constant modulus transmissions over a Gaussian multiple-access channel causes the fusion center in Figure \ref{Fig:problem_setup} to have access to only $y_{L}$, rather than $\{x_{l}\}_{l=1}^{L}$. Clearly, $y_{L}$ has less information about $\theta$ than $\{x_{l}\}_{l=1}^{L}$. In what follows, we quantify this loss by examining the efficiency of the minimum (asymptotic) variance estimator, and comparing it with the benchmark for the centralized problem, $[I(\eta)]^{-1}$, for different distributions on the sensing noise, $\eta$. Using Theorem \ref{lemma:bounds_fisher_info}, it is shown that there is no loss in efficiency if and only if $\eta$ is Gaussian. 

\subsection{The Estimator}
\label{ssec:estimator}

To estimate $\theta$, we normalize $y_{L}$ in (\ref{eqn:yn}) and define:
\begin{equation}
z_{L} \mathop{:=} \frac{y_{L}} {L} = \sqrt{\rho} e^{j \omega \theta}
\frac{1}{L} \sum_{l=1}^{L} e^{j \omega \eta_{l}} + \frac{\nu}{L},
\label{def_zl}
\end{equation}
where $z_{L} = |z_{L}| \exp(j \angle z_{L}) = z_{L}^{R} + j z_{L}^{I}$,
and $z_{L}^{R}$ and $z_{L}^{I}$ are the real and imaginary parts
of $z_{L}$, respectively.
Also $\mathbf{z}_{L} \mathop{:=} [z_{L}^{R} \text{    }
z_{L}^{I}]^{T}$ and $\bar{\mathbf{z}}(\theta) \mathop{:=} 
[E[z_{L}^{R}] \text{
} E[z_{L}^{I}]]^{T} = \sqrt{\rho} [\varphi_{R}(\omega) \cos{\omega
\theta} - \varphi_{I}(\omega) \sin{\omega \theta} \quad \text{    }
\varphi_{R}(\omega) \sin{\omega \theta} + \varphi_{I}(\omega)
\sin{\omega \theta}]^{T}$. 

Given $y_{L}$ (or equivalently $\mathbf{z}_{L}$), the estimator with the smallest asymptotic variance is given by
\cite[(3.6.2), pp. 82]{porat1994}
\begin{equation}
\hat{\theta}_{L} = \argmin_{\theta}  [\mathbf{z}_{L} -
\bar{\mathbf{z}}(\theta)] \boldsymbol{\Sigma}^{-1} (\theta)
[\mathbf{z}_{L} - \bar{\mathbf{z}}(\theta)]^{T},
\label{eqn:est_theta_min_prb}
\end{equation}
where
\begin{equation}
\boldsymbol{\Sigma} (\theta) = \left[ \begin{array}{cc}
\Sigma_{11}(\theta) & \Sigma_{12}(\theta)\\
\Sigma_{21}(\theta) & \Sigma_{22}(\theta) \end{array} \right]
\label{eqn:sigma_matrix_elements}
\end{equation}
is the $2\times 2$ asymptotic covariance matrix of $\mathbf{z}_{L}$,
satisfying $\lim_{L\to\infty} \sqrt{L} [\mathbf{z}_{L} -
\bar{\mathbf{z}}(\theta)] = \mathcal{N} (\mathbf{0},
\boldsymbol{\Sigma} (\theta))$. Its elements are given by
\begin{align*}
\Sigma_{11}(\theta) &= \rho \left[ v_{c} \cos^{2}(\omega \theta) + v_{s} \sin^{2}(\omega \theta) \right] \\
\Sigma_{22}(\theta) &= \rho \left[ v_{s} \cos^{2}(\omega \theta) + v_{c} \sin^{2}(\omega \theta) \right] \\
\Sigma_{12}(\theta) &= \Sigma_{21}(\theta) = \rho (v_{c} - v_{s})
\sin(\omega \theta) \cos(\omega \theta),
\end{align*}
where $v_{c} \mathop{:=} \text{var} [\cos (\omega \eta_{l})] = 1/2 +
\varphi_{R}(2 \omega) /2 - \varphi_{R}^{2}(\omega)$ and $v_{s} \mathop{:=}
\text{var}[\sin (\omega \eta_{l})] = 1/2 - \varphi_{R}(2 \omega)/2 -
\varphi_{I}^{2}(\omega)$.

Estimators of the form in (\ref{eqn:est_theta_min_prb}) have an
asymptotic variance given by \cite[Lemma 3.1]{porat1994}
\begin{equation}
\text{AsV} (\omega) = \left[
\left( \frac{\partial \bar{\mathbf{z}} (\theta)}{\partial \theta} \right)^{T}
\boldsymbol{\Sigma}^{-1} (\theta)  \left( 
\frac{ \partial \bar{\mathbf{z}}(\theta)} {\partial \theta} \right)  \right]^{-1}. \label{eqn:asv_min_mm}
\end{equation}
Substituting $\partial \bar{\mathbf{z}}(\theta)/\partial \theta =
\sqrt{\rho} \omega [ -\varphi_{R}(\omega) \sin{\omega \theta} -
\varphi_{I}(\omega) \cos{\omega \theta} \quad
\varphi_{R}(\omega) \cos{\omega \theta} - \varphi_{I}(\omega)
\sin{\omega \theta} ]^{T}$ and $\boldsymbol{\Sigma}^{-1} (\theta)$
whose elements can be expressed in terms of $\Sigma_{11}(\theta)$,
$\Sigma_{22}(\theta)$ and $\Sigma_{12}(\theta)$, the asymptotic
variance is given by
\begin{align}
\text{AsV} (\omega) &= \frac{2 v_{c}
v_{s}} {\omega^{2} \left[ v_{s} \varphi_{I}^{2} (\omega) + v_{c}
\varphi_{R}^{2} (\omega) \right]} \nonumber \\
&= \frac{ \left( 1 + \varphi_{R}(2 \omega) - 2 \varphi_{R}^{2}(\omega) \right) \left( 1 - \varphi_{R}(2 \omega) -
2 \varphi_{I}^{2}(\omega) \right) } {\omega^{2} \left[ \varphi_{R}^{2} (\omega) \left( 1 + \varphi_{R}(2 \omega) - 2 \varphi_{R}^{2}(\omega) \right) + \varphi_{I}^{2} (\omega) \left( 1 - \varphi_{R}(2 \omega) -
2 \varphi_{I}^{2}(\omega) \right) \right] }. \label{eqn:asv_defn}
\end{align}
Note that $\text{AsV}(\omega)$ depends on the sensing noise through its characteristic function, and does not depend on the channel noise variance, $\sigma_{\nu}^{2}$, which washes out for large $L$. 

\subsection{Asymptotic Efficiency}
\label{ssec:asym_eff}

We now address the asymptotic efficiency of $\hat{\theta}_{L}$ and
characterize the condition under which $\text{AsV}(\omega)$ can be
made arbitrarily close to $[I(\eta)]^{-1}$:
\begin{thm}
The estimator in (\ref{eqn:est_theta_min_prb}) can be arbitrarily
close to being asymptotically efficient by the proper choice of $\omega$, that is,
\begin{equation}
\inf_{\omega \in (0,2\pi/\theta_{R}]} \text{AsV}(\omega) = \frac{1}
{I(\eta)}, \label{eqn:thm_eff_est}
\end{equation}
if and only if $\eta$ is Gaussian. \label{thm:eff_est}
\end{thm}

\begin{IEEEproof}
We begin by showing that if (\ref{eqn:thm_eff_est}) holds, then
$\eta$ is Gaussian. Using Theorem \ref{lemma:bounds_fisher_info}, the inequalities in
(\ref{eqn:lem_ineq_imag}) and (\ref{eqn:lem_ineq_real}) can be
rewritten for $\omega > 0$ as
\begin{align}
\frac{\omega^{2} \varphi_{I}^{2}(\omega)} {\left( \frac{1}{2}
\left[ 1 + \varphi_{R} (\omega) \right] - \varphi_{R}^{2}(\omega)
\right)} &< I(\eta), \label{eqn:thm_ineq_imag_rewrite} \\
\frac{\omega^{2} \varphi_{R}^{2}(\omega)} {\left( \frac{1}{2}
\left[ 1 - \varphi_{R} (\omega) \right] - \varphi_{I}^{2}(\omega)
\right)} &< I(\eta), \label{eqn:thm_ineq_real_rewrite}
\end{align}
where we use that when $\omega \neq 0$, (\ref{eqn:lem_ineq_imag}) and (\ref{eqn:lem_ineq_real}) are strict inequalities.
Adding the inequalities in (\ref{eqn:thm_ineq_imag_rewrite}) and
(\ref{eqn:thm_ineq_real_rewrite}), rearranging the resulting
inequality and recalling (\ref{eqn:asv_defn}), we have
\begin{equation}
\frac{1} {I(\eta)} < \text{AsV}(\omega), \qquad \omega \in
(0,2\pi/\theta_{R}]. \label{eqn:thmproof_fisher_ineq}
\end{equation}
Equation (\ref{eqn:thmproof_fisher_ineq}) indicates that the infimum in (\ref{eqn:thm_eff_est}) is not attained for any non-zero finite value of $\omega$. Since $\omega$ is bounded above, the only way for (\ref{eqn:thm_eff_est}) to hold is when $\lim_{\omega\to 0} \text{AsV} (\omega) = [I(\eta)]^{-1}$. 
It is easy to
verify, using L'Hospital's rule, that $\lim_{\omega \to 0}
\text{AsV} (\omega) = \sigma_{\eta}^{2}$, the variance of
$\eta_{l}$. Therefore, for (\ref{eqn:thm_eff_est}) to hold, we have
$[I(\eta)]^{-1} = \sigma_{\eta}^{2}$. The only distribution that
satisfies this is the Gaussian \cite[Lemma 1.19]{Johnson2004}. This completes the proof of the first half. 

To show that (\ref{eqn:thm_eff_est}) holds when $\eta_{l}$ is Gaussian,
$\varphi(\omega) = e^{-\omega^{2} \sigma_{\eta}^{2} /2}$ is
substituted into (\ref{eqn:asv_defn}) to yield:
\begin{equation}
\text{AsV} (\omega) = \frac{1}{\omega^{2}} e^{-\sigma_{\eta}^{2}
\omega^{2}} \left( e^{2 \sigma_{\eta}^{2} \omega^{2}} - 1 \right)^2,
\label{eqn:asv_min_gauss}
\end{equation}
which is non-decreasing in $\omega$, since
\begin{equation}
\frac{\partial \text{AsV} (\omega)} {\partial \omega} = \frac{2 e^{-2 \sigma_{\eta}^{2} \omega^{2}}} {\omega^{3}} \left( e^{2\sigma_{\eta}^{2} \omega^{2}} - 1\right) \left[ (1 - e^{2 \sigma_{\eta}^{2} \omega^{2}}) + 2\sigma_{\eta}^{2} \omega^{2} + 2 \sigma_{\eta}^{2} \omega^{2} e^{2 \sigma_{\eta}^{2} \omega^{2}} \right] \geq 0,
\label{eqn:thmproof_derivative_asv}
\end{equation}
for $\omega>0$. 
\end{IEEEproof}

The phase modulated scheme considered here has the advantage of constant modulus transmissions. 
Due to the use of phase modulation, the result in Theorem \ref{thm:eff_est} is related to the efficiency of the estimator of
a location parameter using the empirical characteristic function (ECF), defined as $\hat{\varphi} (\omega) \mathop{:=} L^{-1} \sum_{l=1}^{L} e^{j \omega x_{l}}$. It can be seen from (\ref{def_zl}) that $z_{L} = \sqrt{\rho} e^{j \omega \theta} \hat{\varphi} (\omega) + \nu/L$ is related to the ECF through scaling and additive noise. The
efficiency of empirical characteristic function based estimators has
been considered for arbitrary parameters (that is, not just location parameters) in \cite{Feuerverger1981}, but with a continuum of infinitely
many values of the argument, $\omega$, of the ECF. In
the current distributed estimation application, the evaluation of $\hat{\varphi}(\omega)$ for many values of $\omega$ at the fusion center
corresponds to many transmissions per sensor observation, requiring large bandwidth. In contrast, we
consider a single value of $\omega$ for estimation, requiring a single transmission per sensor. The analog
transmissions are assumed to be appropriately pulse-shaped and phase
modulated to consume finite bandwidth.

When the sensing noise distribution is symmetric, the 
cost function on the right hand side of
(\ref{eqn:est_theta_min_prb}) that needs to be minimized can be
expressed as
\begin{align}
c(\theta) =& [\mathbf{z}_{L} - \bar{\mathbf{z}}(\theta)]
\boldsymbol{\Sigma}^{-1} (\theta) [\mathbf{z}_{L} -
\bar{\mathbf{z}}(\theta)]^{T} \nonumber\\
=& \frac{1} {2 \rho^{2} v_{c} v_{s}}  \Big[ -4 \rho^{3/2}
v_{s} \varphi(\omega) [z_{L}^{I} \sin(\omega \theta) + z_{L}^{R}
\cos(\omega \theta)] + 2 \rho^{2} v_{s} \varphi^{2}(\omega)
\nonumber\\
&+ \rho (v_{c}-v_{s}) \left( (z_{L}^{I})^{2} -
(z_{L}^{R})^{2} \right) \cos(2 \omega \theta) - 2\rho (v_{c} +
v_{s}) z_{L}^{I}z_{L}^{R} \sin(2 \omega \theta) \nonumber\\
&+ \rho (v_{c}+v_{s}) \left( (z_{L}^{I})^{2} + (z_{L}^{R})^{2}
\right) \Big]. \label{eqn:cost_fn_expression}
\end{align}
Differentiating with respect to $\theta$, we have
\begin{align}
\frac{\partial c(\theta)}{\partial \theta} =& \frac{2 \omega z_{L}^{R} \cos(\omega \theta)}{\rho v_{c} v_{s}}  \left[ \frac{z_{L}^{I}}{z_{L}^{R}} - \tan (\omega \theta) \right] \nonumber \\
& \times \left[ \left( 1 + \frac{z_{L}^{I}}{z_{L}^{R}} \tan (\omega \theta) \right) v_{c} + \left( 1 - \frac{\sqrt{\rho} \varphi (\omega)}{z_{L}^{R} \cos (\omega \theta)} \frac{z_{L}^{I}}{z_{L}^{R}} \tan (\omega \theta) \right) v_{s} \right].
\label{eqn:cost_fn_der}
\end{align}
The values of $\theta$ at which (\ref{eqn:cost_fn_der}) is zero are given by
\begin{equation}
\theta \in \left\{ \frac{n \pi \pm \frac{\pi}{2}} {\omega}, \frac{1}{\omega} \angle z_{L}, \frac{ \angle z_{L} + 2 n \pi \pm \frac{\pi}{2} } {\omega} \right\},
\label{eqn:c_th_diff_zero}
\end{equation}
where $\omega \neq 0$ and $n \in \mathbb{Z^{+}}$. The value of $\theta$ that minimizes $c(\theta)$ is easily verified by substituting the values of $\theta$ from (\ref{eqn:c_th_diff_zero}) into (\ref{eqn:cost_fn_expression}) and is given by
\begin{equation}
\hat{\theta} = \frac{1}{\omega} \angle z_{L}.
\label{eqn:est_theta_min_prob_soln}
\end{equation}
Hence, in the presence of symmetric noise, the estimator in (\ref{eqn:est_theta_min_prb}) that minimizes the asymptotic variance reduces to the simple expression in (\ref{eqn:est_theta_min_prob_soln}), which was first considered in \cite{robust_est_cihan_adarsh_icassp}. However, in \cite{robust_est_cihan_adarsh_icassp}, neither the optimality (in terms of minimizing the asymptotic variance) nor the efficiency of the estimator in (\ref{eqn:est_theta_min_prob_soln}) was considered. 

\subsection{Quantifying Relative Efficiency}
\label{ssec:quant_eff}

One way of interpreting Theorem \ref{thm:eff_est} is to observe that when the sensing noise is Gaussian, no information is lost by analog
phase modulation if $\omega$ is chosen sufficiently small. On the other hand, information is lost when the
sensing noise follows other distributions. To see this more clearly, we define the relative efficiency between the asymptotic variance and the Fisher information as: 
\begin{equation}
\mathcal{E} (\eta) = \left[ I(\eta) \inf_{\omega\in(0,2\pi/\theta_{R}]} \text{AsV} (\omega) \right]^{-1}.
\label{eqn:defn_efficiency}
\end{equation}
It can easily be verified that $\mathcal{E} (\eta)$ is scale-invariant in the sense that $\mathcal{E} (\alpha \eta)=\mathcal{E} (\eta)$ for any $\alpha\in \mathbb{R}$. Moreover, based on Theorem \ref{thm:eff_est} and (\ref{eqn:thmproof_fisher_ineq}), $0\leq \mathcal{E} (\eta) \leq 1$, where the equality in the upper-bound is achieved only if $\eta$ is Gaussian. 

The relative efficiency in (\ref{eqn:defn_efficiency}) depends only on the distribution of the
sensing noise. The values of $\mathcal{E}(\eta)$ for several
distributions are provided in Table \ref{table:info_eff_values}. The result in Table \ref{table:info_eff_values} for the Gaussian case has been established in Theorem \ref{thm:eff_est}. For the Laplace sensing noise, $\varphi(\omega) = (1 + \omega^{2} \sigma_{\eta}^{2}/2)^{-1}$, $\text{AsV} (\omega) = \sigma_{\eta}^{2} (1 + \sigma_{\eta}^{2} \omega^{2}/2)/(1 + 2 \sigma_{\eta}^{2} \omega^{2})$, and $\inf_{\omega\in(0,2\pi/\theta_{R}]} \text{AsV} (\omega) = 3\sigma_{\eta}^{2}/4$, by inspecting the third derivative of $\text{AsV}(\omega)$. Similarly for the Cauchy case, $\varphi(\omega) = e^{-\gamma \omega}$, $\text{AsV}(\omega) = e^{2\gamma\omega} (1-e^{-2\gamma\omega})/2\omega^{2}$, and $\inf_{\omega\in(0,2\pi/\theta_{R}]} \text{AsV} (\omega) = 4 \gamma^{2} e^{c} (1-e^{-2c})/c^{2}$, by examining the first derivative of $\text{AsV}(\omega)$ where $\gamma$ is the scale parameter of the Cauchy random variable, $c \mathop{:=} 2 + W(-2e^{-2})$, and $W(\cdot)$ is the Lambert $W$-function \cite{Lambert1996}. For the uniform distribution, an extension of the definition in (\ref{eqn:fisher_info_defn}) can be used to argue that the Fisher information is infinite \cite[pp. 119]{Lehmann1998}, and the relative efficiency of the estimator as defined in (\ref{eqn:defn_efficiency}) is zero. 

\begin{table}[tb]
\begin{centering}
\begin{tabular}{|c||c|c|c|c|}
\hline Distribution & Gaussian & Laplace & Cauchy & Uniform\\
\hline $\mathcal{E}(\eta)$ & $1$ & $2/3$ & $0.5 c^{2} e^{-c} (1 - e^{-c})^{-1} \approx 0.65$ & $0$\\
\hline
\end{tabular}
 \caption{$\mathcal{E}(\eta)$ for different distributions.}
 \label{table:info_eff_values}
\end{centering}
\end{table}

We have seen that the Gaussian sensing noise is the only distribution with the highest possible efficiency when the observations $x_{l}$ are transmitted with phase modulation over Gaussian multiple-access channels and the estimator in (\ref{eqn:est_theta_min_prb}) is used. However, it is possible that other sensing noise distributions, which yield less efficiency, have better asymptotic variances. This is because efficiency is defined relative to the Fisher information. For example, for Laplace sensing noise, the proposed estimator is not asymptotically efficient, but has better asymptotic variance than in the Gaussian case, since its inverse Fisher information, $[I(\eta)]^{-1}$, is lower. In conclusion, Gaussian sensing noise has the only distribution that does not suffer a loss in efficiency when the sensed data $x_{l}$ is mapped to constant modulus transmissions over Gaussian multiple-access channels. 

\section{Numerical Results}
\label{sec:simulations}

\begin{figure}[tb]
\begin{minipage}[b]{1.0\linewidth}
  \centering
  \centerline{\epsfig{figure=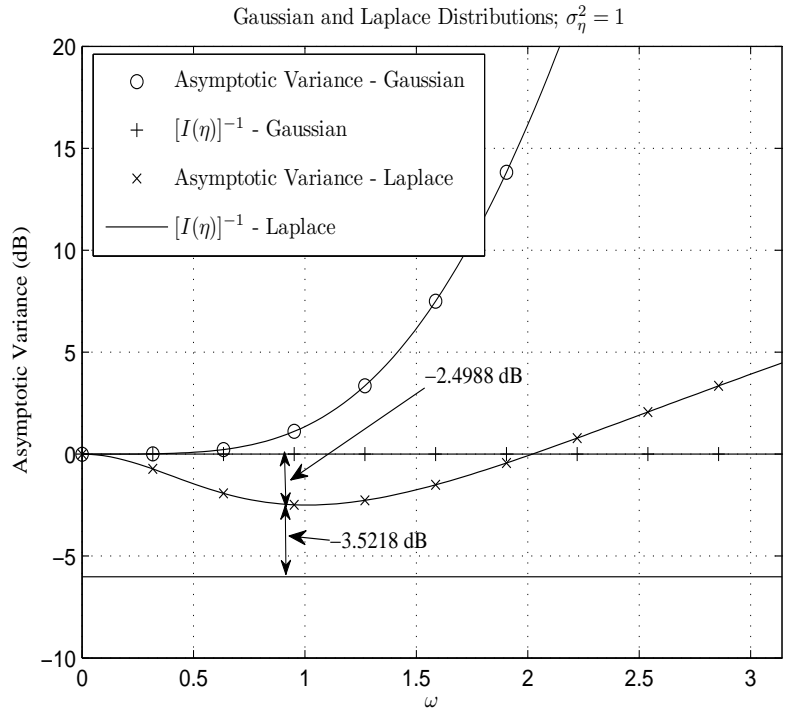,width=12cm,height=10cm}}
\end{minipage}
\caption{Plot of asymptotic variance vs. $\omega$.}
\label{Fig:gauss_laplace_asv_crlb}
\end{figure}

\begin{figure}[tb]
\begin{minipage}[b]{1.0\linewidth}
  \centering
  \centerline{\epsfig{figure=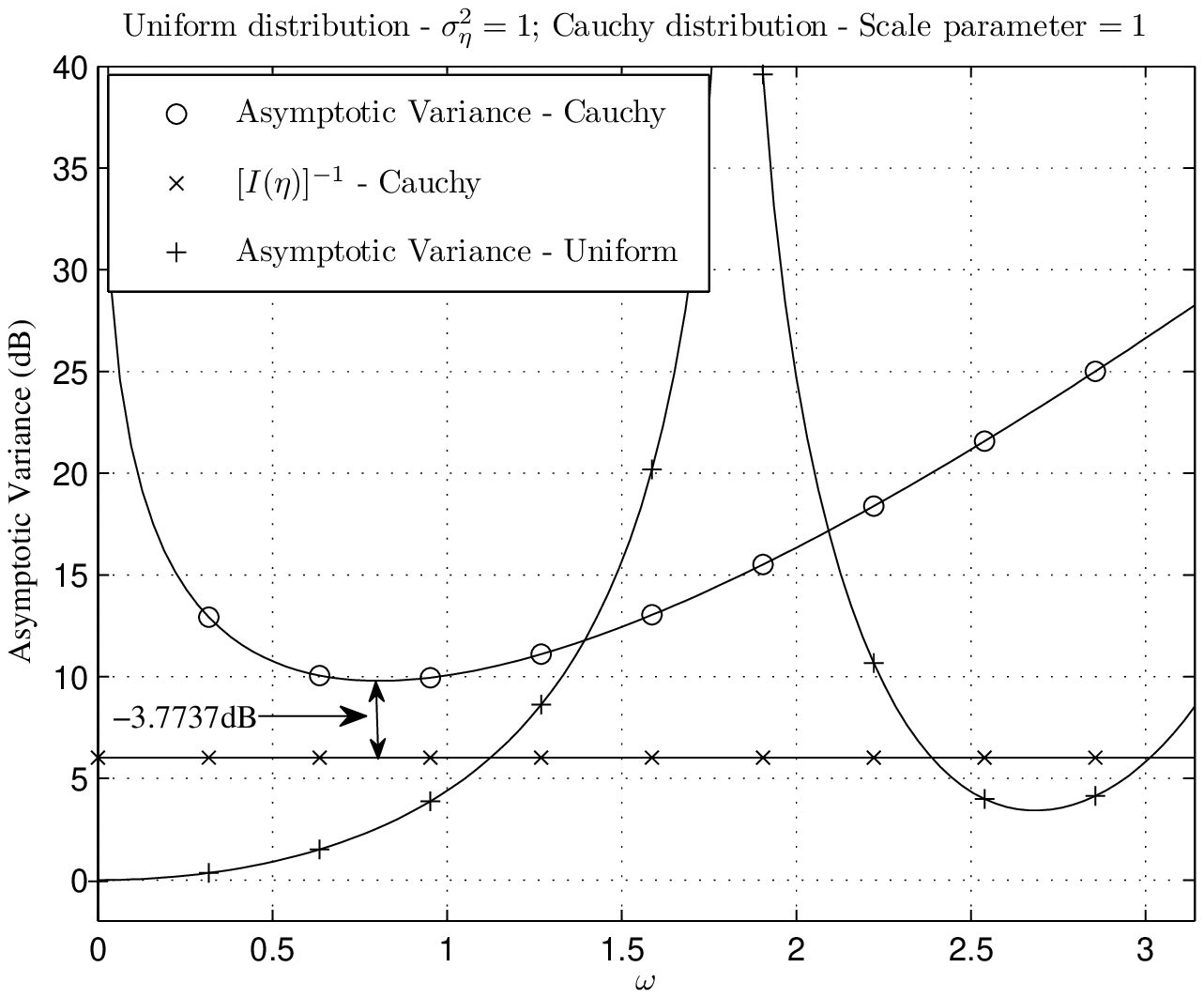,width=12cm,height=10cm}}
\end{minipage}
\caption{Plot of asymptotic variance vs. $\omega$. Note that the value of $[I(\eta)]^{-1}$ is 0 ($-\infty$ dB) for the uniform sensing noise case and is not shown.}
\label{Fig:uniform_cauchy_asv_crlb}
\end{figure}

In Figures \ref{Fig:gauss_laplace_asv_crlb} and
\ref{Fig:uniform_cauchy_asv_crlb}, the asymptotic variance and the
value of $[I(\eta)]^{-1}$ in dB are plotted versus $\omega$, when the sensing noise is Gaussian, Laplace, uniform and
Cauchy distributed. 

From Figure \ref{Fig:gauss_laplace_asv_crlb}, it can be seen that the asymptotic variance
approaches $[I(\eta)]^{-1}$ only as $\omega\to0$ for
Gaussian sensing noise, and is bounded away from $[I(\eta)]^{-1}$ for other values of $\omega$. 
The estimator in (\ref{eqn:est_theta_min_prb}) is not efficient when the sensing noise is non-Gaussian. 
Using the definition of relative efficiency in (\ref{eqn:defn_efficiency}), it can seen from Figure \ref{Fig:gauss_laplace_asv_crlb} that $\mathcal{E} (\eta)$ in the case of Gaussian sensing noise is $0$dB, and in the case of Laplace sensing noise is about $-3.5$dB. In Figure \ref{Fig:gauss_laplace_asv_crlb}, it can be verified that $\inf_{\omega} \text{AsV} (\omega) \approx 0.75$, which is about $-2.5$dB at $\omega=1/\sqrt{2}$, which is lower than the Gaussian sensing noise case. 

From Figure \ref{Fig:uniform_cauchy_asv_crlb} the relative efficiency
for the Cauchy noise case is about $-3.8$dB, verifying the value shown in
Table \ref{table:info_eff_values}. The inverse Fisher information for the uniform case is 0 ($-\infty$ dB) and is not shown in Figure \ref{Fig:uniform_cauchy_asv_crlb}. The relative efficiency as defined in (\ref{eqn:defn_efficiency}), for uniform noise, is therefore zero. When the sensing noise follows the Cauchy, uniform or Laplace distributions, the estimator is not asymptotically efficient. 

\section{Conclusions}
\label{sec:conclusions}

In this paper, we considered the relationship between the Fisher
information and the characteristic function through two bounds. The condition for equality
was also derived, for the first time in literature.

This result was used to prove the asymptotic efficiency of a distributed estimator that minimizes the asymptotic variance
in the presence of Gaussian sensing noise. In all cases, the loss in
efficiency was quantified through a scale-invariant relative efficiency metric that takes
values between 0 and 1. This metric depends only on the distribution
of the sensing noise used, and was computed for the Gaussian, Laplace, Cauchy and uniform cases. These relative efficiency values can be interpreted as the amount of information lost due to constant modulus transmissions over Gaussian multiple-access channels relative to having perfect access to all sensor measurements. 
Numerical evaluations confirm the
result that the estimator in (\ref{eqn:est_theta_min_prb}) is asymptotically efficient only when the sensing noise is Gaussian.

\bibliography{robustest}
\end{document}